\documentclass[journal=jacsat,manuscript=article]{achemso}

\usepackage{chemformula} 
\usepackage[T1]{fontenc} 

\usepackage{mathptmx}      
\usepackage{mhchem}

\usepackage{subfig}
\usepackage{siunitx}
\usepackage{booktabs}
\usepackage[english]{babel}
\usepackage{xcolor}
\usepackage{multirow}
\usepackage{nameref}
\usepackage{textcomp}
\usepackage{float}
\SectionNumbersOn 

\author{Cassia Lux}
\author{Sabrina Kerz}
\affiliation[Klitzing]
{Soft Matter at Interfaces, Department of Physics, 64289 Darmstadt, Germany}
\author{Catarina C. Ribeiro}
\affiliation[Stark]
{Physics of Surfaces, Department of Material Science, 64287 Darmstadt, Germany}
\author{Jennifer Bareuther}
\affiliation[Rehahn]
{Macromolecular Chemistry: Chemistry of Polymers, Department of Chemistry, 64287 Darmstadt, Germany}
\author{Johannes Lützenkirchen}
\affiliation[KIT]
{Institute for Nuclear Disposal, Karlsruhe Institute of Technology, 76021 Karlsruhe, Germany}
\author{Sebastian Stock}
\affiliation[Klitzing]
{Soft Matter at Interfaces, Department of Physics, 64289 Darmstadt, Germany}
\author{Michaelis Tsintsaris}
\affiliation[Klitzing]
{Soft Matter at Interfaces, Department of Physics, 64289 Darmstadt, Germany}
\author{Matthias Rehahn}
\affiliation[Rehahn]
{Macromolecular Chemistry: Chemistry of Polymers, Department of Chemistry, 64287 Darmstadt, Germany}
\author{Robert W. Stark}
\affiliation[Stark]
{Physics of Surfaces, Department of Material Science, 64287 Darmstadt, Germany}
\author{Regine von Klitzing}
\affiliation[Klitzing]
{Soft Matter at Interfaces, Department of Physics, 64289 Darmstadt, Germany}
\email{klitzing@smi.tu-darmstadt.de}
\phone{+49(0)6151-16-24506}
\affiliation[Unknown University]
{Soft Matter at Interfaces, Department of Physics, 64289 Darmstadt, Germany}

\title[An \textsf{achemso} demo]
  {Conceptualizing flexible papers using cellulose model surfaces and polymer particles}

\keywords{cellulose model surface; core-shell particles; microgels; force mapping}

\begin{document}



\begin{abstract}
\noindent Cellulose, as a naturally abundant and biocompatible material, is still gaining interest due to its high potential for functionalization. This makes cellulose a promising candidate for replacing plastics. Understanding how cellulose interacts with various additives is crucial for creating composite materials with diverse properties, as it is the case for plastics. In addition, the mechanical properties of the composite materials are assumed to be related to the mobility of the additives against the cellulose. Using a well-defined cellulose model surface (CMS), we aim to understand the adsorption and desorption of two polymeric particles (core-shell particles and microgels) to/from the cellulose surface. The nanomechanics of particles and CMS are quantified by indentation measurements with an atomic force microscope (AFM). AFM topography measurements quantified particle adsorption and desorption on the CMS, while peak force AFM measurements determined the force needed to move individual particles. Both particles and the CMS exhibited pH-dependent charge behavior, allowing a tunable interaction between them. Particle adsorption was irreversible and driven by electrostatic forces. In contrast, desorption and particle mobility forces are dominated by structural morphology. In addition, we found that an annealing procedure consisting of swelling/drying cycles significantly increased the adhesion strength of both particles. Using the data, we achieve a deeper understanding of the interaction of cellulose with polymeric particles, with the potential to advance the development of functional materials and contribute to various fields, including smart packaging, sensors, and biomedical applications.
\end{abstract}

\section{Introduction}
\label{intro}

Cellulose, as the most abundant natural polymer present on Earth, has attracted significant attention as a potential substitute for synthetic polymers in the development of functionalized paper and cellulose-based materials.\cite{Wang2021,Thomas2018,Gilbert2017} In its unmodified form, paper has a high biocompatibility, recyclability, and biodegradability \cite{Rinaudo2008a}; however, it also has a low mechanical stability \cite{Strand2017}. The primary forces responsible for the stability of paper, composed of a network of cellulose fibers, include van der Waals forces and hydrogen bonds. These bonds are susceptible to breaking when the fibers come into contact with water and swell, as the distance between the two fiber surfaces increases.\cite{Dunlop-Jones1991} To enhance the stability of paper, functional additives are added during the papermaking process. The additives reinforce the interconnections between cellulose fibers through physical or covalent bonding.\cite{Lindstrom2005} Depending on the additive a multitude of specialized properties of paper is achievable. Commonly, additives are used to enforce mechanical stability. Moreover, innovative paper properties can be induced, such as conductivity in case of paper electronics\cite{Agate2018}, substance uptake for drug delivery applications\cite{Khine2020}, responsiveness for the use of biosensors\cite{Ratajczak2020} or even capacity for energy storage\cite{Wang2021a}.\\

Planar cellulose model surfaces (CMSs) are a good and valid approach for the investigation of paper functionalization with typically used additives as they offer a lot of advantages.\cite{Ahola2008,Kontturi2006,Gunnars2002}  First of all, a CMS is a well-defined controllable substrate, allowing a reproducible in-depth study of different functionalization techniques and disentangling the different surface effects responsible for the interactions. It can also be considered as a simplified system of the complex material paper. Both the paper network and the fiber itself are highly structured systems, which also have a high content of other components such as lignin and hemicelluloses. Finally, CMSs, which are thin and smooth films, allow the use of many techniques for surface characterization such as Atomic Force Microscopy (AFM), Ellipsometry and reflectometry methods.\cite{Kontturi2006}\\

The preparation of CMS requires either a solution \cite{Aulin2009} or a stable and homogeneous suspension of nanometer sized cellulose particles\cite{Edgar2003}. Cellulose, due to its complex fiber structure and the presence of intra- and intermolecular hydrogen bonds and hydrophobic interactions\cite{Medronho2012}, is not dissolvable in water and most organic solvents, with the exception of two-component solvents such as \textit{N,N}-dimethylacetamide (DMA)/LiCl and ionic liquids.\cite{Medronho2012} However, the use of such solvent systems leads to the integration of a solid non-volatile component in the CMS, such as LiCl in the previously mentioned solvent system. The subsequent removal of this component requires an additional washing step, which may change the morphology of the model surface. Various established techniques such as spin-coating, dip-coating, spray-coating, Langmuir-Blodgett (LB) deposition, and Langmuir-Schäfer (LS) deposition methods are applied for the preparation of model surfaces.\cite{Kontturi2019,Kontturi2006} Leporatti et al. studied the interaction of a CMS with a cellulose colloidal sphere in the presence of cationic copolymers to mimic the interaction with wet-strength agents in paper. The used CMS with a very low roughness was prepared by spin-coating cellulose dissolved in \textit{N}-methylmorpholine-N-oxide.\cite{Leporatti2005}\\

In addition to studying the interactions with typically used additives in the paper industry, CMSs are also used in the development of new materials. The used additives in the cellulose composites impart new properties or functionalities in paper materials and can range from low molecular substances such as surfactants\cite{Penfold2007} and metal ions\cite{Bethke2018} over high molecular polymers, e.g. chitosan\cite{Strnad2023}, to (polymeric) nanoparticles. A large interest also exists in combining cellulose fibers with stimuli-responsive polymers such as poly(\textit{N}-Isopropylacrylamide) (PNIPAM) or  poly(2-(dimethylamino) ethyl methacrylate) (PDMAEMA).\cite{Li2019} Studies characterizing the interaction of these different types of additives aim to understand the effect on the properties of the resulting paper material, and how the preparation parameters and the surface forces affect the functionalization. Therefore, the central question is which interaction forces occur between the cellulose surface and the additive. The overarching goal is the disentanglement of different effects on the interaction stemming from the additive properties such as electrostatics, deformability, and mobility of the polymer particles.\\

\begin{figure*}[]
                 \centering
                 \includegraphics[scale=0.8]{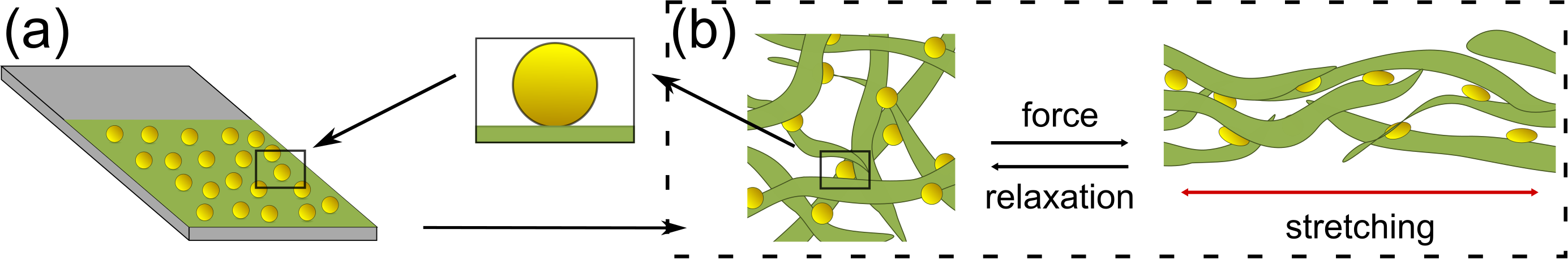}
                 \caption{Scheme of (a) the experimental model system consisting of a cellulose model surface and polymeric particles to gain an understanding of the interactions in the composite materials and how it affects the resulting mechanical properties. (b) The polymeric particles can be used in so-called flexible papers which show a high flexibility when a mechanical force is applied.}
                 \label{SketchElasticPaper}
\end{figure*}

In this study, we investigated the interaction between a CMS and two different elastomeric particles to understand the paper functionalization with polymeric nanoparticles. In comparison to classical linear polymers, the particles do not enter the fiber but are deposited onto the surface of the fiber. Additionally, the retaining of the spherical shape and the larger dimensions allow new properties of paper, such as increasing the flexibility and elasticity of paper. The flexibility of paper can be enhanced by adding the particles during the papermaking process, leading to an accumulation between the cellulose fibers. When a mechanical force is applied to the prepared paper, it results in the deformation and shearing of the elastomeric particles, as well as the parallel alignment of the cellulose fibers (Figure \ref{SketchElasticPaper}). Thereby, the naturally rigid and brittle paper gains flexibility. Such materials have the potential to substitute e.g. synthetic rubbers.\\

The elastic particles used in this study are core-shell particles (CSPs) and poly(N-isopropylacrylamide) (PNIPAM) microgels (MGs). The CSP have a hard polymethyl methacrylate (PMMA) core and a soft polyethylene acrylate (PEA) shell, both highly crosslinked. The MGs are low crosslinked hydrogels with a higher crosslinked core due to one-batch synthesis. The MG exhbibits a higher deformability and dangling ends at its surface compared to the harder CSP with its smooth surface. Both types of particles and CMS show pH dependent surface charge, which allows pH adjustment for optimum adsorption. The impact of the pH on the charge of the particles and the different particles structures on the adsorption process on a CMS were studied. For both composites a pH regime is found in which the particles are positively charged and the CMS is negatively charged. We use atomic force microscopy (AFM) to analyze the distribution of the adsorbed particles, their adhesion to the CMS, and their deformability. These measurements provide insights into the characteristics and behavior of the adsorbed particles within the composite system and, at last, in their suitability as particles for elastic papers.

\section{Materials and methods}
\label{sec:exppart}

\subsection{Materials}
\label{materials}
\subsubsection{CMS and pH solutions:} Microcrystalline cellulose (\SI{20}{\micro m}), \textit{N,N}-Dimethylacetamid (DMA, anhydrous, \SI{99.8}{\percent}), hydrochlorid acid (\SI{1}{mol/l}), sodium hydroxide solution (\SI{1}{mol/l}), and lithium chloride (\SI{99}{\percent}) were purchased from Merck KGaA (Darmstadt, Germany). Sulfuric acid (\SI{96}{\percent}) and hydrogen peroxide (\SI{30}{\percent}) were purchased from Carl Roth (Karlsruhe, Germany). Ultrapure water was obtained from a Milli-Q-system from Merck with a resistance of \si{18} {M$\Omega$}\,cm.

\subsubsection{MG synthesis:} \textit{N}-isopropylacrylamide (NIPAM) and \textit{N},\textit{N}'-methylenbisacrylamid (BIS) were purchased from Merck KGaA (Darmstadt, Germany).
2,2'-azobis-2-methyl-propan-imidamide dihydrochloride (AAPH) was purchased from Cayman Chemical Company (Cayman Chemical, USA).

\subsubsection{CSP synthesis:} The monomers ethylacrylate (EA), methyl methacrylate (MMA), allyl methyacrylate (ALMA), glycidyl methacrylate (GlyMA) were obtained from ACROS Organics (Thermo Fisher Scientific, Massachusetts, USA). The monomers were destabilised by basic alumium oxide (also obtained from ACROS Organics). The surfactant Disponil\textregistered A 1080 was obtained from BASF (Ludwigshafen, Germany). Sodium disulfate and sodium persulfate was obtained Merck KGaA (Darmstadt, Germany).

\subsection{Preparation and synthesis}
\label{prepSynth}

\subsubsection{Preparation of the cellulose model surface (CMS)}
\label{prepCMS}
The solution of cellulose was prepared under an inert argon atmosphere using a synthesis protocol described in literature.\cite{Carapeto2017} The microcrystalline cellulose and LiCl were first dried in a vacuum oven at \SI{100}{\celsius} for \SI{2}{h}. \SI{0.5}{g} microcrystalline cellulose and \SI{20}{\milli\litre} DMA were placed into a flask and heated to \SI{150}{\celsius} under stirring. After obtaining a homogeneous suspension, the mixture was cooled down to \SI{100}{\celsius}. \SI{1.5}{g} LiCl was added and the mixture was stirred for \SI{10}{min}. The suspension was then slowly cooled down to room temperature over a period of \SI{2}{h} by decreasing the heating temperature stepwise. This resulted in a highly viscous, yellow, but clear cellulose solution, which was stored in the fridge until further use.\\

To prepare the model surfaces, double side polished silicon wafers from Siegert Wafer (Aachen, Germany) were first cut to sizes of 1.5\,x\,\SI{1.5}{\centi\meter\squared} and rinsed with ultra-pure water. They were etched in a 3:1 (\ce{H2SO4}:\ce{H2O2}) piranha solution for \SI{30}{min} and rinsed thoroughly with ultra-pure water. The etched silicon wafers were spin-coated with the cellulose solution at \SI{3000}{RPM} for \SI{30}{s} with an acceleration of \SI{1000}{RPM/s}. To anneal the films, the spin-coated silicon wafers were heated at \SI{180}{\celsius} for \SI{5}{min}. The LiCl was then rinsed out of the CMS by placing the CMS wafer into ultra-pure water for \SI{5}{min}.
 
\subsubsection{Synthesis of CSP}
\label{synthCSP}
The polymerization of the PMMA-PEA CSPs was carried out in a \SI{1}{\litre} double-walled reactor under nitrogen atmosphere, with attached reflux condenser, wing stirrer, and the aid of a thermostat (\SI{90}{\celsius} setting). First, the template consisting of \SI{3.6}{g} seed monomer MMA, \SI{0.4}{g} crosslinker ALMA, \SI{0.052}{g} surfactant Disponil, and \SI{280}{g} degassed water was weighed and mixed in a \SI{500}{ml} flask. For the initiator system two times \SI{0.05}{g} sodium disulfonate and one time \SI{0.3}{g} sodium persulfonate were each weighed into snap cap vials and dissolved in \SI{5}{ml} water each. The template solution was poured into the pre-tempered reactor, and the initiator was added immediately. The mixture was stirred for \SI{30}{min} (turbidity occurs after approx. \SI{20}{min}). The monomer emulsion consisting of \SI{84}{g} shell monomer EA, \SI{9.3}{g} crosslinker ALMA, \SI{0.53}{g} surfactant, \SI{0.48}{g} potassium hydroxide, and \SI{109.3}{g} degassed water was weighed into another flask and homogenized using an ultrasonic finger. Then, the monomer emulsion was continuously added to the reactor using a reciprocating pump at \SI{80}{\celsius}. The particle diameter was controlled by regularly taking small samples via the outlet and analyzing them with a zeta sizer. After the complete addition of the monomer emulsion, the dispersion was stirred in the reactor for another hour before it was drained.\\

To prepare the elastomeric CSP with a cationic surface charge, following the method described by Winter et al. \cite{Winter2021}, the previously prepared particles were reinitiated. \SI{74.3}{g} of the particle dispersion was transferred to a pre-tempered \SI{250}{ml} double-walled reactor under nitrogen atmosphere. The reactor was equipped with a reflux condenser, a wing stirrer, and a thermostat (\SI{90}{\celsius} setting). To start reinitiation, an initiator system consisting of two times \SI{0.012}{g} sodium disulfate and \SI{0.138}{g} sodium persulfate was added to the dispersion. The mixture was stirred for \SI{30}{min}. A second monomer emulsion was prepared, consisting of \SI{0.3}{g} EA, \SI{0.7}{g} GlyMA, \SI{0.036}{g} surfactant Disponil, and \SI{1.16}{g}. The monomer emulsion was continuously added to the particle dispersion and the dispersion was further stirred for \SI{1.5}{h}. A sample of the dispersion was taken and analyzed using PREUSSMANN reagent (4-(4-nitrobenzyl) pyridine). To perform the analysis, two spatula tips of PREUSSMANN reagent was dissolved in \SI{5}{ml} isopropanol and 3-4~drops ammonia. Then, about 1-\SI{2}{ml} of the emulsion sample was added to the mixture. The sample was stirred for several hours, and the color progression was observed. The remaining emulsion was drained through a filter into a three-neck flask equipped with a stirring fish and a reflux condenser. The dispersion was further heated to approximately \SI{90}{\celsius} using an oil bath. A third monomer emulsion, consisting of \SI{0.76}{g} methylamine, \SI{0.34}{g} potassium hydroxide, and \SI{3}{g} degassed water, was added dropwise to the dispersion using a syringe. The reaction mixture was stirred for several days. The resulting emulsion was transferred into a polypropylene bottle and stored for further use.

\subsubsection{Synthesis of MG}
\label{synthMG}
Microgel particles (MGs) were synthesized using a well-established precipitation polymerization reaction protocol described in literature.\cite{Witte2019,Sanson2010,Pelton1986,Pelton2000} The reaction was carried out in a double-walled glass reactor under an inert nitrogen atmosphere. \SI{2.15}{g} of the monomer NIPAM and \SI{0.15}{g} (\SI{5}{mol\%}) of the cross-linker BIS were dissolved in \SI{120}{ml} of water. The solution was degassed with constant stirring (\SI{1000}{RPM}) by passing a continuous nitrogen flow AAPH initiator dissolved in \SI{1}{\milli\litre} of water via a syringe. The reaction was carried out for \SI{90}{min} at \SI{80}{\celsius} and \SI{1000}{RPM}. The obtained MGs were cleaned by dialysis for a minimum of \SI{10}{days} (10 cycles with \SI{120}{ml} dispersion against \SI{5}{\litre} water per cycle). Furthermore, the MGs were purified by removing smaller oligomers through sedimentation during centrifugation (\SI{10000}{\cdot g}, \SI{30}{min}), followed by redispersion. This purification process was repeated at least four times. The obtained MGs with a hydrodynamic diameter of \SI{620}{nm} (measured by dynamic light scattering, LS instruments, Fribourg, Switzerland) were dried by lyophilization and stored in a freezer until further use.

\subsection{Composite films}
\label{method:composite}
Composite films using the CMSs and the particles were prepared by dip coating. The CMSs were placed into the particle solutions with set pH for \SI{15}{min} and afterwards into ultra pure water with the same pH for \SI{5}{min}. They were dried at air. For the annealed composite films, the composite films underwent 10 cycles of swelling and drying by placing the composite film into ultra pure water set to pH\,3 for \SI{30}{min} and then leaving it to completely dry at air for several hours. 

\subsection{Characterization methods}
\label{CharMethod}

\subsubsection{Zeta potential measurements}
\label{method:zeta}
The electrophoretic mobility $U_\text{E}$ was measured with a Zeta Sizer Nano (Malvern Panalytical, Malvern, UK)  at a concentration of \SI{0.006}{wt\%}. The zeta potential $\zeta$ was calculated by

\begin{eqnarray}
\zeta = \frac{U_\text{E} \cdot 3 \eta}{2 \epsilon\cdot f(\kappa a)} \label{eq:Henry}
\end{eqnarray}

with the dynamic viscosity $\eta$ and the dielectric constant $\epsilon$. $f(\kappa a)$ is the Henry function and can be approximated to $\frac{3}{2}$ using the Smoluchowski approximation as the ratio of particle diameter and the Debye length is $\gg1$. 

\subsubsection{Streaming potential measurements}
\label{method:streaming}
The zeta potential of the CMS was determined via streaming-current measurements using the SurPass set-up (Anton Paar, Graz, Austria). Prior to the measurements, the CMSs are spin-coated on 1\,x\,2\,cm silicon wafers. Two  samples were mounted to the sample holders of the adjustable gap cell so that the CMS would be exposed towards the flow channel. Subsequently, the sample holders with the samples were inserted into the adjustable gap cell and fixed in the set-up with pressure sensors and electrodes connected. The gap between the two surfaces was adjusted to about \SI{100}{\micro\meter}. As a last step before the experiment, a flow check was carried out to make sure that the pressure/flow rate was linear and identical for flow in both directions. 

Experiments were carried out in ultra-pure water. The water (500 ml reservoir) was kept under Argon atmosphere at room temperature during the measurements. Automated titrations to vary the pH were carried out by adding \SI{10}{mM} HCl to the reservoir.
 
For each pH, the system was flushed with the current solution for 15 minutes before starting the streaming current measurement. For each pH condition, six distinct measurements were carried out (three from left to right and three from right to left). Experimental errors as estimated from the six measurements were typically below \SI{1}{mV} (standard deviation), and for the starting pH or the very low pH amounted to about \SI{2}{mV}. 

\subsubsection{AFM}
\label{method:afm}

Atomic force microscopy (AFM) was used to characterize the topography of the composites, and for the indentation and the pushing particles measurements. The AFM topography and indentation measurements were carried out with the MFP-3D SA (Asylum Research, Oxford Instruments, California, USA) in tapping mode. 

\paragraph*{Topography}
The topography measurements were done at ambient conditions in air and the cantilevers used are the AC160TS-R3 with a silicon probe and a tip diameter of 7-\SI{8}{nm}, also from Asylum Research. The obtained images are plane-fitted and first order fitted with regards to the underlying substrate.

\paragraph*{Indentation measurements}
The indentation measurements were carried out in water, set to a specific pH, with the HQ:NSC19/Cr-Au cantilevers (for CSP, obtained from MikroMasch, NanoAndMore GmbH, Germany) or the BL-AC40TS (for MG, obtained from Olympus, Germany). In the first part, the trigger points were fixed to \SI{10}{nN} for the CSPs and \SI{2}{nN} for the MGs. The obtained force curves were fitted with the Hertz model, 

\begin{eqnarray}
F&=&\frac{4}{3} \cdot \frac{E}{1 - m^2}\cdot (R \cdot d^3 )^{\frac{1}{2}}  \label{eq:Ind}
\end{eqnarray}

with the force of indentation $F$, the elastic modulus $E$, the Poisson radius $m$, the probe radius $R$ and indentation depth $d$. 

For the indentation measurements at higher trigger points and higher indentation, only a section of the whole force curve is fitted. The whole force curve is split into 10 equal sections and fitted with Equation \ref{eq:IndOff}, which includes an offset $d_0$ in the indentation. By using the offset, we are able to calculate the elastic moduli (\textit{E} moduli) at higher indentation independent of the moduli at lower indentations.

\begin{eqnarray}
F_{\text{offset}}&=&\frac{4}{3} \cdot \frac{E}{1 - m^2}\cdot (R \cdot (d-d_0)^3 )^{\frac{1}{2}} \label{eq:IndOff}
\end{eqnarray}

\begin{figure}[]
                 \centering
                 \includegraphics[width = 8 cm]{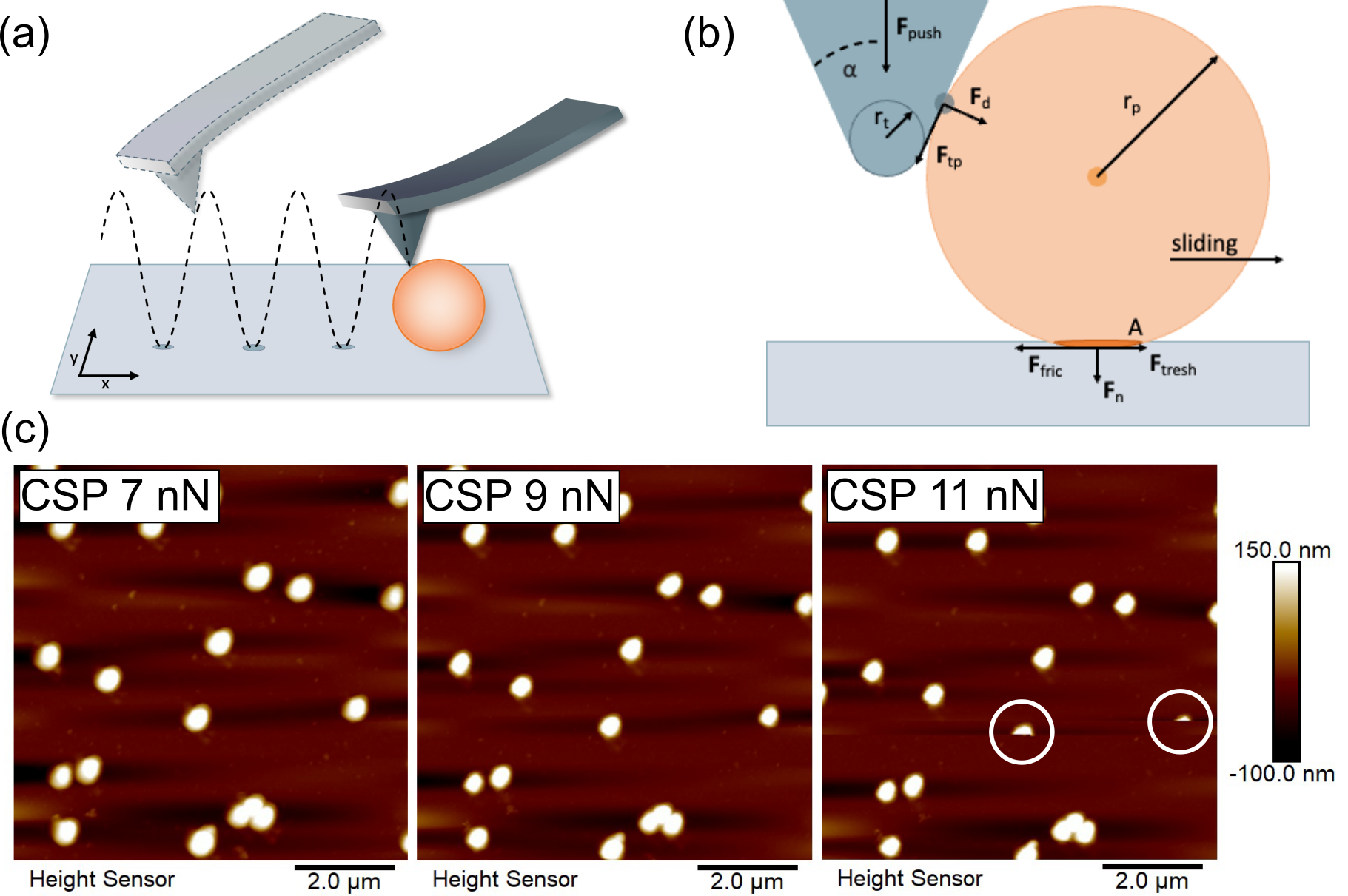}
                 \caption{(a) Scheme of the pushing experiment using peak force tapping mode, illustrating the sinusoidal trajectory of the tip while scanning the sample surface. (b) Scheme of interacting forces acting between the tip and a particle.  Images adapted from \cite{Schiwek2015}. (c) Exemplary images of the AFM tapping force measurements for the CSP on the CMS. All AFM images have a size of 8\,x\,\SI{8}{\micro m} and a height scale of \SI{250}{nm}. The force for the scanning is incrementally increased and the displacement of the particles observed. For the CSP at pH\,3 (on silicon wafer) \SI{7}{nN} and \SI{9}{nN} is not sufficient to move the particles, but at \SI{11}{nN} the displacement of two particles is observed.}
                 \label{fig:methodpushpar}
\end{figure}

For both cases an exemplary curve with its fit is given in the SI in Figure S1.
\paragraph*{Pushing particles}
\label{method:PushPar}
To characterize the pH dependency of the adsorption behavior of the particles on the substrates, a pushing experiment was carried out using AFM. The cantilever tip was utilized as both a pushing manipulator and as a force sensor to investigate the forces necessary to push the particles on the substrate following a procedure proposed by Schiwek et al.\cite{Schiwek2015} The sample surface was imaged using the peak force tapping mode. The cantilever tip oscillates far from its resonance frequency and the feedback loop controls the instantaneous deflection that corresponds to the desired maximum applied force. Therefore, the cantilever motion follows a sinusoidal trajectory as represented in Figure \ref{fig:methodpushpar}a. It should be noted that the applied force on the substrate/particles is a normal/vertical force, not a tangential force. The maximum applied force after each consecutive scan is incrementally increased until the particles begin to move. This enables an analysis of the force needed to overcome the adhesion and friction forces between the particles and the substrate. 

During the pushing process, the particle is subjected to both adhesive and frictional forces. The adhesive force depends on the contact area between the particle and the substrate and is related to the interfacial energy of the materials in contact. The frictional force is a measure of the resistance to motion of the particle on the substrate and is also related to the contact area. The particles begin to slide once the force $F_\text{push}$ applied by the tip overcomes the adhesion and friction forces between the particle and the substrate, as illustrated in Figure \ref{fig:methodpushpar}b). A detailed description of contact mechanics between a particle and the cantilever tip can be found in the work of Schiwek et al.\cite{Schiwek2015}

The vertical force applied by the tip on the particle $F_\text{push}$ can be split into a tangential force component $F_\text{tp}$ and a component pointing to the center of the particle $F_\text{d}$. $F_\text{tp}$ is mostly dissipated in friction between the tip and the particle, whereas $F_\text{d}$ gives rise to a vertical force $F_\text{n}$ and a horizontal threshold force $F_\text{thresh}$. The opening angle of the tip $\alpha$ can be used to determine the trigonometric ratio of the force components. $F_\text{fric}$ causes resistance of the particle to motion, once $F_\text{thresh} \geq F_\text{fric}$ the particle begins to slide. The $F_\text{push}$ can be correlated to $F_\text{tresh}$ with
\begin{eqnarray}
F_\text{thresh}=F_\text{push}\cdot\sin \alpha \cdot\cos \alpha.
\end{eqnarray}

The pushing experiments were realized in the PeakForce tapping mode with a Dimension Icon AFM using a ScanAsyst-fluid+ cantilever (\SI{0.8}{N/m}), both from Bruker AXS (Santa Barbara, CA). The cantilever had a spring constant of \SI{0.8}{N/m} and a resonance frequency of \SI{150}{kHz}, which was deduced from the thermal noise method.\cite{Butt1995} A scan rate of \SI{0.5}{Hz}, frequency of \SI{2}{kHz} and an amplitude of \SI{100}{nm} were set. All measurements were performed in water with set pH to study the pH-dependency of the adsorption behaviour of the particles on the substrate.

\section{Results and discussion}
\label{results}

\subsection{Characterization of CSPs and MGs}
\begin{figure*}[]
                 \centering
                 \includegraphics[width = 10 cm]{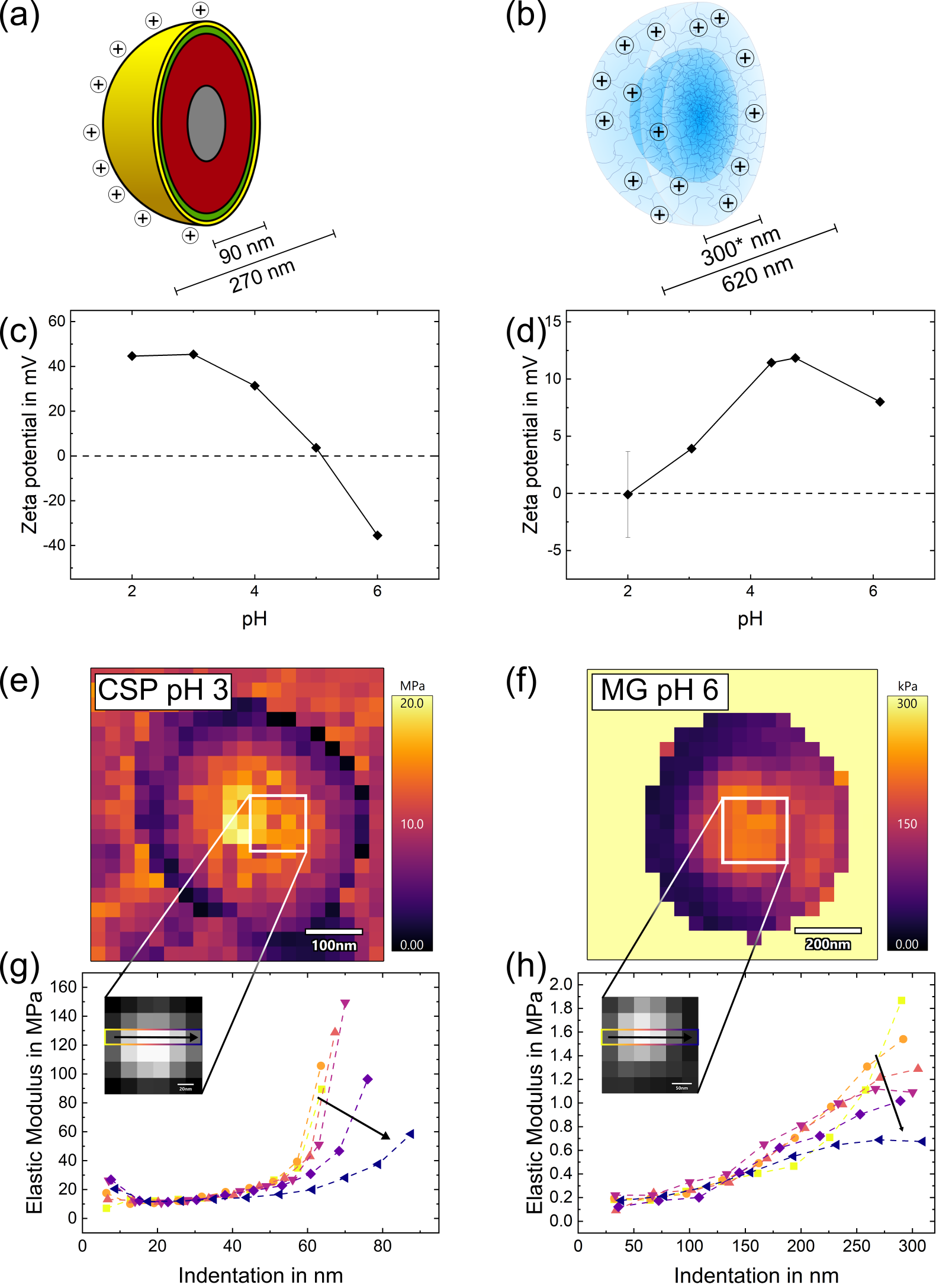}
                 \caption{The CSPs and MGs characterized by AFM and zeta potential measurements. The scheme of the structure of the CSP (a) and the MG (b) is given on the basis of the synthesis and known literature. The CSP has the core-shell structure. The MG has a typical porous structure with a higher crosslinked core and a fluffy shell. Both particles exhibit a pH dependent positive charge as shown by the zetapotential measurements (c,d). By indentation measurements the \textit{E} moduli and the structure of the CSP on CMS (e) and the MG on Si (f) are characterized. For the CSP the indentation depth was approximately \SI{10}{nm}, for the MG \SI{50}{nm}. The measurement was repeated with a higher indentation depth to resolve the inner structure of the particles (g,h) and the force curve stepwise fitted to give \textit{E} moduli in dependence of the indentation depth. For each particle a cross section of six force curves measured at the center of the particles are depicted.}
                 \label{characterizationParticles}
\end{figure*}

Figure \ref{characterizationParticles}a,b displays a schematic of the CSP and the MG and their dimensions. The CSP consists of a 90 nm diameter PMMA core, coated with a softer PEA shell. The PEA shell has a thickness of approximately \SI{90}{nm}, with a thin layer of PGlyMA, functionalized with an amine on its surface. The hydrodynamic diameter is measured to be about \SI{270}{nm} via DLS. The net charge density is obtained by electrophoretic mobility measurements and plotted in dependence of the pH in Figure \ref{characterizationParticles}c. The amine groups at the surface of the CSP lead to a positive charge, increasing with decreasing pH. The maximum potential is \SI{40}{mV} at pH\,2. At pH\,5, the CSP are uncharged. Increasing the pH to pH\,6, leads to a negative charge. We could show that the observed negative charge is not due to a decomposing effect of CSP, as the effect is reversible when decreasing the pH again. Literature previously discussed the specific adsorption of hydroxide ions at polymeric interfaces leading to measured negative zeta potentials.\cite{Zimmermann2010} The other polymers present in the CSP do not possess ionizable groups and additionally would not be detected by the zeta sizer due to the high crosslinking in the shell.\\

As shown in Figure \ref{characterizationParticles}b, the PNIPAM MG also has a core-shell structure, but it is less defined compared to the CSP as the crosslinking fraction gradually decreases towards the outside. This is due to the faster reaction kinetics of the crosslinker BIS, leading to an accumulation of the crosslinker at the center of the MGs. This has been shown in previous studies by our group by AFM\cite{Witte2021,Backes2018,Burmistrova2011}, by TEM measurements of nanoparticles incorporated MGs\cite{Witt2019} and measuring reaction kinetics\cite{Forg2022}, and in literature by reaction kinetics\cite{Wu1994} and using small-angle neutron scattering\cite{Saunders2004,Stieger2004}. The MG's outer shell is commonly described as fluffy, primarily composed of single chains and low crosslinker content. A positive charge of the MGs is induced by the protonation of the polymerization initiator AAPH, in detail, by the amidine group. The initiator distribution within the MG is still the subject of ongoing research. Previous publications have suggested that the initiator is located on the MG's surface, measured by small-angle neutron scattering\cite{Scotti2016}. For our initial interpretation, we assume that the initiator is uniformly distributed throughout the microgel, and the ionizable (and charged) groups are located in the outer region of the core and within the fluffy shell. The pH dependency on the zeta potential was measured by electrophoretic mobility measurements (Figure \ref{characterizationParticles}d). It is important to mention that the obtained zeta potential should be considered with precaution due to the MG's porous structure. As the porous structure would allow a permeability, the force/current applied on the MGs would be significantly smaller independent of the charge. However, due to the amount of crosslinker and resulting small mesh sizes, the porosity of the MGs is not high enough to allow a permeability and the electrophoretic mobility, and therefore the zeta potential, can be correlated directly to the charge density. The zeta potential measurements show a positive charge of \SI{10}{mV} at pH\,6. Decreasing the pH first leads to a small increase in charge. Below pH\,4 the zeta potential decreases until pH\,2, at which the MGs no longer appear to be charged. The decreasing of the charge density with decreasing pH can be attributed to an increase in ionic strength leading to a shielding off of the charges.\\

\begin{figure*}[]
                 \centering
                 \includegraphics[width = 12 cm]{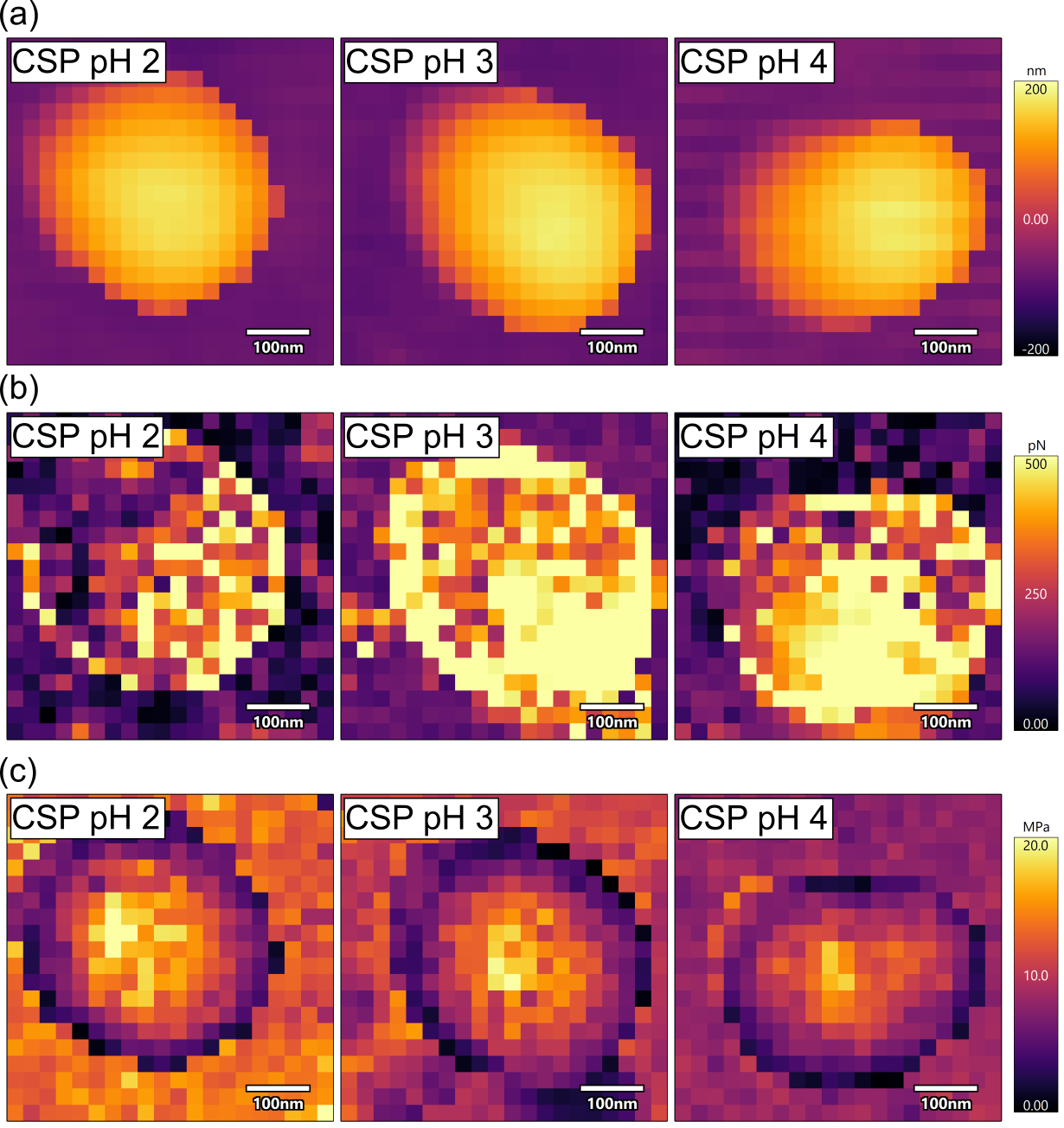}
                 \caption{Indentation measurements of the CSPs at varying pH, with the (a) height, (b) cantilever adhesion, and (c) \textit{E} moduli maps (calculated by the Hertz model). The measurements were carried in water set to the corresponding pH with the NSC19-Au cantilever. All images have a lateral scale of \SI{500}{nm}.}
                 \label{EModCSP}
\end{figure*}

\begin{figure}[]
                 \centering
                 \includegraphics[scale=0.25]{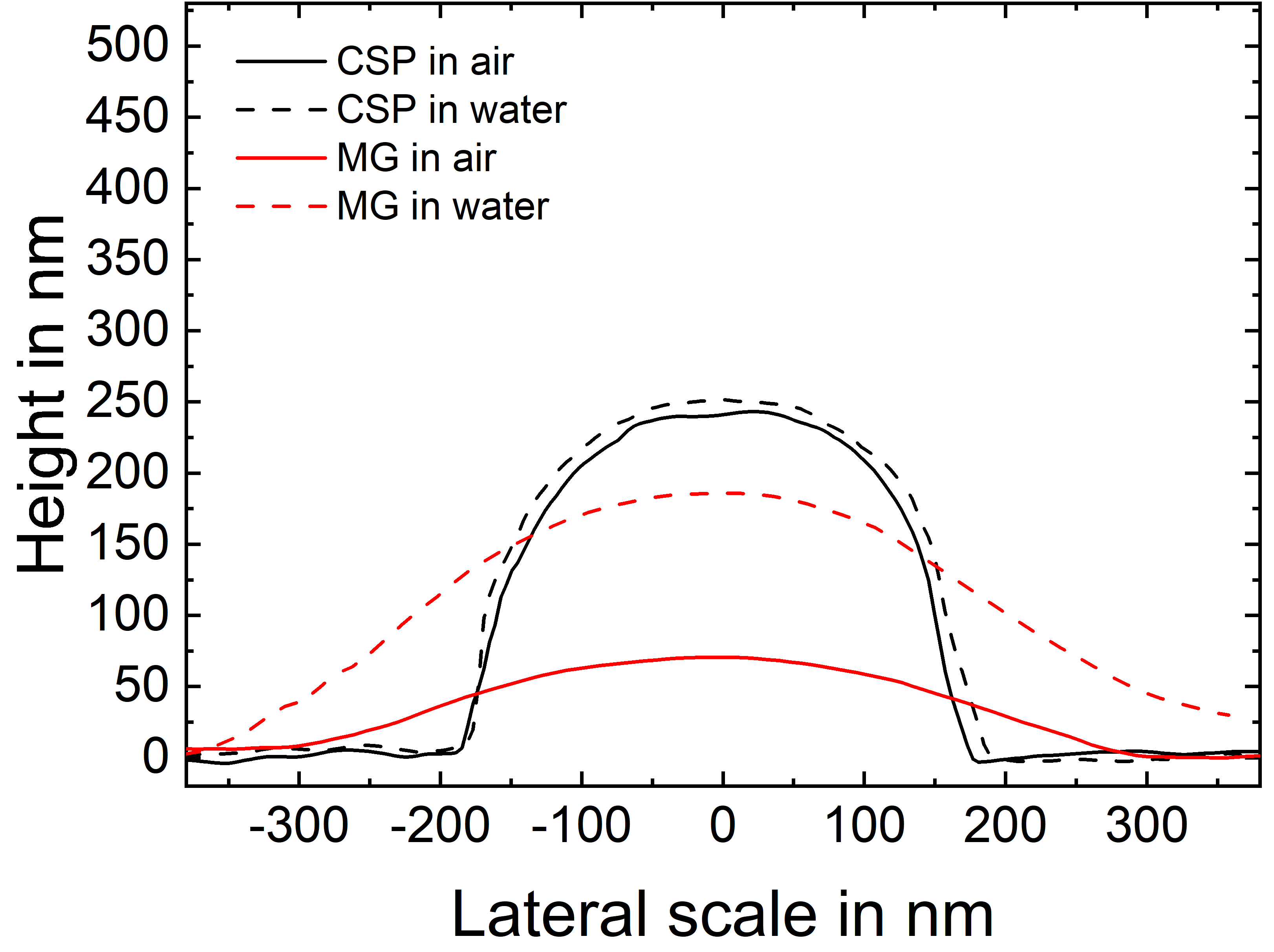}
                 \caption{Cross-section of CSP and MG in water and in air measured by AFM.}
                 \label{Crosssection}
\end{figure}

To resolve the structures of the CSPs, force indentation measurements were carried out on adsorbed particles. In Figure \ref{characterizationParticles}c the lateral map of the \textit{E} moduli at different positions on the CSP is shown exemplary for pH\,3. The height-, adhesion (to cantilever)\footnote{Please be aware that one has to distinguish between the adhesion between tip and particle (here) and the adhesion between particle and CMS (later, in peak force measurements)}, and \textit{E} moduli maps at varying pH  are shown in Figure \ref{EModCSP}). The spherical particles are cleary discernible in every map. The \textit{E} moduli are the highest at the center of the particle with \textit{E} moduli up to \SI{20}{MPa}. The \textit{E} moduli decrease outwardly, and at the border of the CSP \textit{E} moduli below \SI{1}{MPa} are measured. The low \textit{E} moduli measured at the particle borders can be attributed to the cantilever slipping caused by the overhanging of the CSP. The cross-section analysis of the CSP (Figure \ref{Crosssection}) indicates that the CSPs mostly maintain their shape. Comparing the \textit{E} moduli profiles of the CSPs at the different pH (Figure \ref{EModCSP}) does not show any significant pH-dependent effect, as the pH-responsive shell only makes up a few nanometers of the whole CSP. The \textit{E} moduli of the CSPs are of the same magnitude as those of the CMS on which the CSP are adsorbed, with the CSP exhibiting \textit{E} moduli of 10\,-\,\SI{20}{MPa} and the CMS 18\,-\,\SI{10}{MPa}.\\

\begin{figure*}[]
                 \centering
                 \includegraphics[scale=0.8]{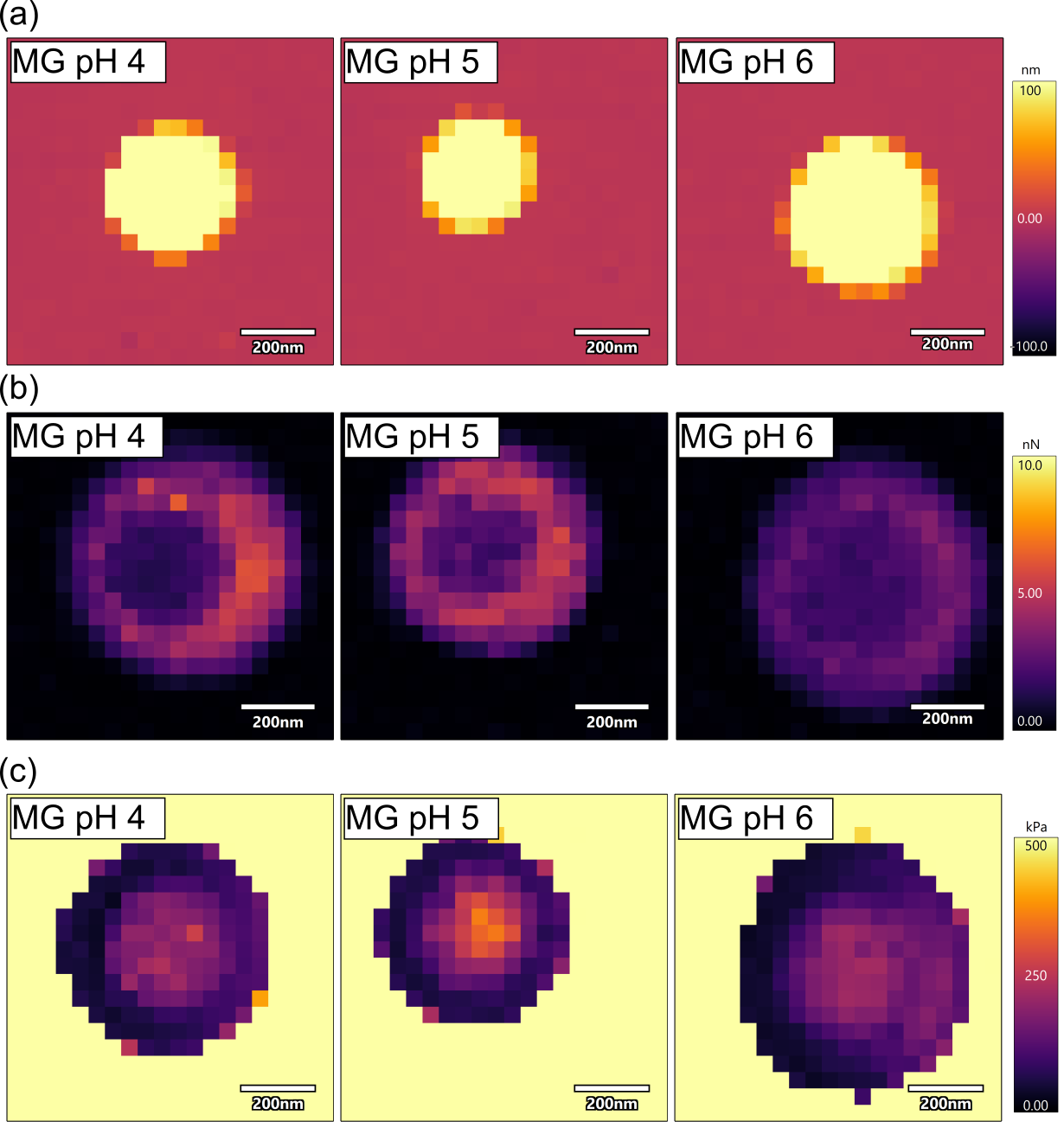}
                 \caption{Indentation measurements of the MGs at varying pH, with the (a) height, (b) cantilever adhesion, and (c) \textit{E} moduli maps (calculated by the Hertz model). The measurements were carried out in water set to the corresponding pH with the BLAC40-TS cantilever. All images have a lateral scale of \SI{1}{\micro m}.}
                 \label{EModMG}
\end{figure*}

As the MGs desorbed from the CMS even with low trigger points, the indentation measurements were carried out for MGs adsorbed on an etched silicon wafer (Figure \ref{EModMG}). The comparison of all maps depicts the particles in the height map laterally smaller than in the adhesion and force maps. The particles have diameters of approx. 200\,-\,\SI{300}{nm} in the height images and approx. 400\,-\,\SI{500}{nm} in the adhesion and force maps (comparison: \SI{620}{nm} hydrodynamic diameter). All images show a spherical shape of the particles. No significant effect of the pH can be observed in the average values. As shown exemplarily for MG adsorbed at pH\,6, the obtained \textit{E} moduli of the MGs are significantly lower than for the CSPs and the substrate, with values around \SI{200}{kPa}. The MGs exhibit a higher \textit{E} modulus at the center due to the higher crosslinker density. The very low \textit{E} moduli at the border of the MGs, due to a slipping off of the cantilever during the measurement, such as it was the case for the CSPs cannot be observed. The MGs deform to maximize the interface, as shown in Figure \ref{Crosssection} due to their loose polymer network, low \textit{E} moduli, and high water content.\cite{Schmidt2010,Backes2017a,Schulte2019} The fluffy shell and dangling ends\cite{Stieger2004,Kyrey2019} of the MGs arrange themselves around the relatively higher crosslinked core, which is also shown in the higher cantilever adhesion around the core (Figure \ref{EModMG}b).\\

To resolve the inner structure of the particles, the indentation measurements were carried out with increasing trigger point at the center of the particle. The higher trigger point leads to a deeper indentation. With increasing indentation depth, the different regimes of stiffness of the particles due to increasing crosslinker density was resolved. The measurements for the CSP particle were carried out at pH\,3. In Figure \ref{characterizationParticles}g, the \textit{E} modulus of the CSP is plotted in dependence of the indentation depth. For sake of clarity, only six curves are plotted and the designating position of the measured curve on the particle is shown in the overlay. The whole force map and all 36 curves can be found in the supporting information (Figure S3), both arranged analog to the force map and a superposition of all curves. The six measured spots exhibit different curve progressions. Curves more located at the center of the particle (orange, pink) exhibit an exponential increase of the \textit{E} moduli with indentation depth, meaning higher \textit{E} moduli are measured at the core of the CSP. Curves further apart from the center of the particle (blue) exhibit a significantly lower exponential increase. This trend is observable for all the curves of the force map (Figure S3).\\

As for the CSPs, an indentation measurement is carried out for a MG adsorbed on a silicon wafer at pH\,6 with increasing trigger points. The \textit{E} modulus is plotted in dependence of the indentation depth in Figure \ref{characterizationParticles}h.  The \textit{E} modulus increases linearly with the indentation depth for all force curves (only 6 shown here, for all force curves see Figure S4), with a maximum \textit{E} modulus of \SI{1.5}{MPa}. For all force curves, an indentation above \SI{250}{nm} is reached before the MG detaches from the substrate. The \textit{E} modulus increases slightly with indentation, which is a bit more pronounced for positions close to the center. In total the effects of indentation depth and distance from the center are much less pronounced than for the CSPs.

\subsection{Characterization of CMS}

\begin{figure*}[]
                 \centering
                 \includegraphics[scale=0.8]{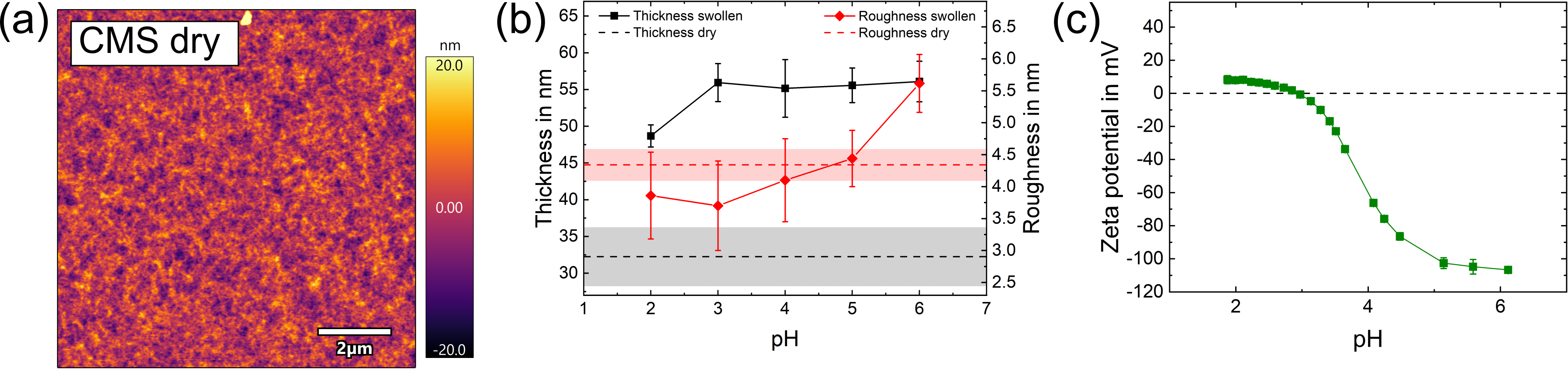}
                 \caption{Characterization of the CMS was carried out with AFM and streaming potential measurements. The topography of the CMS in ambient conditions (a) was measured and the thickness and roughness of the CMS at varying pH characterized (b). The streaming potential measurement gives the pH dependent charge behavior of the CMS (c).}
                 \label{CharacterizationCMS}
\end{figure*}

The prepared CMS is characterized with regards to charge density and morphology in dependence of the pH. As shown by topography AFM measurements (Figure \ref{CharacterizationCMS}a), the surface of the dry CMS measured at ambient conditions is smooth and homogeneous. On a small lateral scale a structuring with low roughness is observed. For the measurement of the thickness, part of the CMS is removed by canula, producing sharp borders, and the height is measured by topography AFM measurements. A thickness of \SI{32}{nm} and roughness of \SI{4.3}{nm} was measured. When in contact with water, the CMS swells. The dependency of the pH on the thickness and the roughness of the CMS is plotted in Figure \ref{CharacterizationCMS}b. The swelling at pH\,6 leads to increase in thickness to \SI{57}{nm}. When decreasing the pH, the thickness remains constant until pH\,3. At pH\,2, the thickness slightly decreases to \SI{46}{nm}. The decrease in swelling from pH\,4 to pH\,2 leads to a stiffening of the CMS as visible in Figure \ref{EModCSP}c from an average \textit{E} modulus of \SI{10}{MPa} to \SI{18}{MPa}. The roughness increases due to charge induced swelling from \SI{3.9}{nm} at pH\,2 to \SI{5.6}{nm} at pH\,6. While the thickness of the swollen CMS is significantly higher than for the dry CMS, the roughness of the dry CMS is in the same range as for the swollen CMS between pH\,2 and pH\,5.\\

Measured by streaming potential measurements, the charge of the CMS in dependence of the pH is depicted in Figure \ref{CharacterizationCMS}c. The naturally occurring carboxy groups in the cellulose lead to an electrokinetic potential at higher pH, up to \SI{-100}{mV}. With decreasing pH and increased protonation of the carboxy groups, the zeta potential decreases sigmoidally. Around pH\,3 the CMS exhibits no charge. At pH\,2 the CMS exhibits a small positive charge, which is attributed to the specific hydronium adsorption at low pH.\cite{Zimmermann2010}

\subsection{Characterization of composites}
\begin{figure*}[]
                 \centering
                 \includegraphics[width = 12 cm]{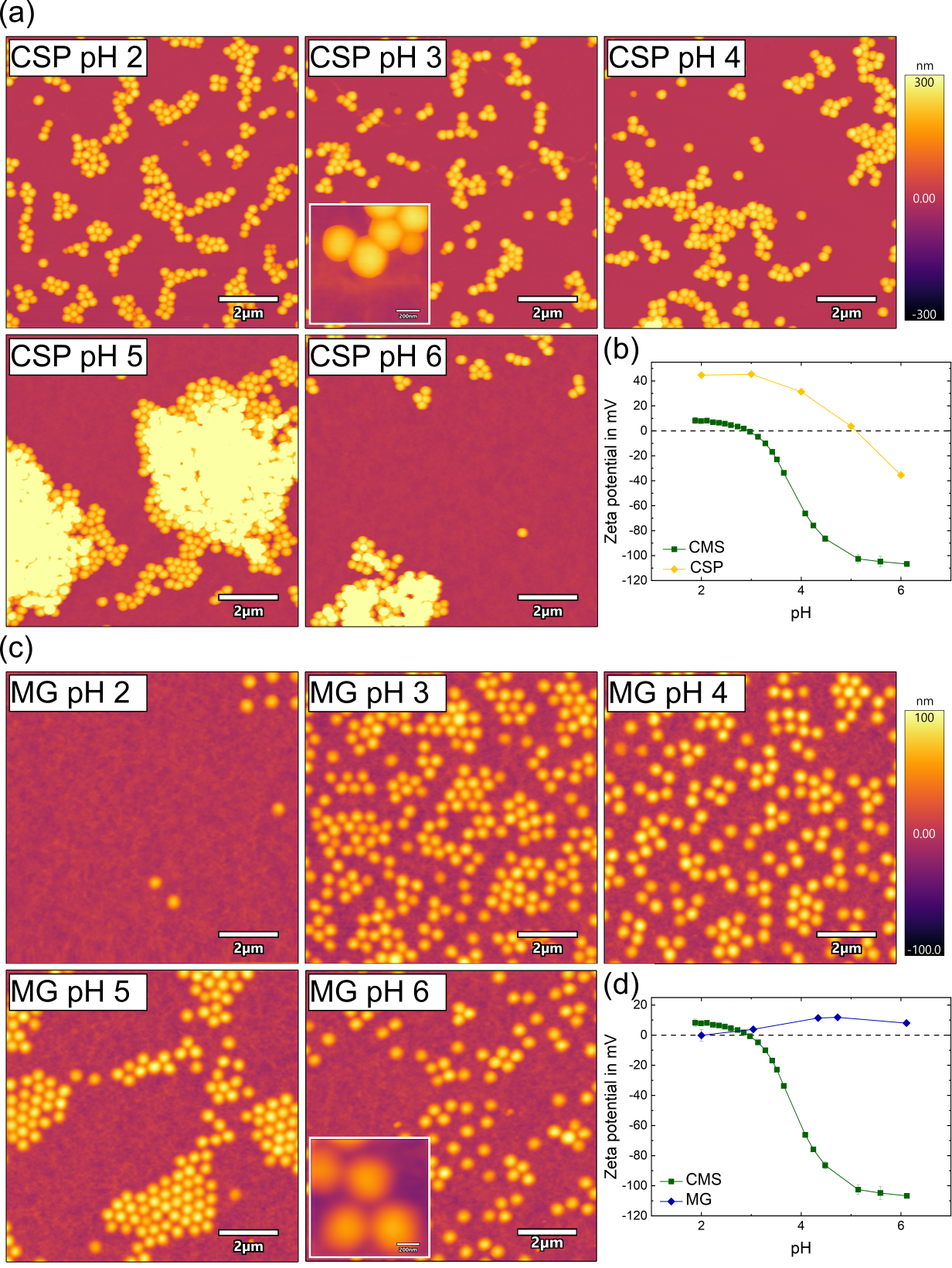}
                 \caption{The adsorption of the core-shell particles (CSP) and microgels (MG) on the cellulose model surface (CMS) at varying pH characterized by AFM measurements (a,c). All AFM images have a 10\,x\,\SI{10}{\micro m} size and the height scale is \SI{600}{nm} for the CSP and \SI{200}{nm} for the MG composites. For the CSP at pH\,3 and a MG at pH\,6 a magnified image is inset with a size of 1\,x\,\SI{1}{\micro m}. The zeta potential measurements of the CSP (b), the MG (d) are compared with the streaming potential measurements of the CMS.}
                 \label{ScansCSPMG}
\end{figure*}

The topography images of the composites prepared from dip-coating of the CMS with the CSPs at different pH are depicted in Figure \ref{ScansCSPMG}a. At pH\,2 to pH\,4, the particles adsorb on the CMS in small areas of hexagonal packing. A reason for the close packed monolayered areas might be capillary forces during the dipping process. The stable adsorption is explained by a relatively high charge for the positively charged CSP and a low negative charge of the CMS (Figure \ref{ScansCSPMG}b). Increasing the pH to pH\,5 leads to the formation of aggregates of the CSPs. Compared to the composites at lower pH, the amount of adsorbed particles is locally significantly higher due to the local multilayer formation and lateral higher packing. This strong aggregation is explained by the low charge of the CSPs. Further increasing the pH to pH\,6 significantly decreases the adsorbed amount of CSP, and the particles cover only a small part of the CMS. The zeta and streaming potential measurements show that at pH\,6 both CMS and CSP (due to the previously mentioned specific hydroxide adsorption) are negatively charged (Figure \ref{ScansCSPMG}b).\\

The topography images of the composite films prepared from dip-coating of the CMS with the PNIPAM MGs at different pH are depicted in Figure \ref{ScansCSPMG}c. At pH\,3 and pH\,4, the particles adsorb in an even distribution with a high particle number density. In this range the MG are slightly charged positively while the CMS exhibits low negative charge (Figure \ref{ScansCSPMG}d), leading to an electrostatic attraction but only a small repulsive force in between the MGs. At pH\,2, only small amounts of particles adsorb. The MGs appear uncharged and the CMS is also uncharged with a low positive charge due to a specific hydronium adsorption possible, resulting in no electrostatic interaction between the MGs and the CMS. At higher pH, pH\,5 and pH\,6, the particles number density is slightly smaller than at pH\,4. Additionally, a high agglomeration with large free spots can be observed for the composite films prepared at pH\,5. This and the overall not clear trend between electrostatic interaction, particle number density and agglomeration is likely caused by desorption during the dipping process and capillary effects during the drying of the composites. The progression of the zeta potential in dependence of the pH also suggest that other effects such as the salt concentration (due to pH setting) impact the adsorption of the MGs. To our best knowledge, little research studying the charge behavior of pure NIPAM MGs (without comonomer) in dependence of pH value and salt concentration exists.\\

The results obtained from the composite scans and zeta potential measurements suggest that the adsorption process is mostly driven by the electrostatic attraction between the charged particles and the CMS. When both the particles and CMS are charged an adsorption with a high particle number density can be observed. It can be derived from the structural composition of the particles and their adsorption behavior that the CSPs are more hydrophobic compared to the MGs. In the absence of charge, the CSPs tend to form larger aggregates. Even at the highest CSP charge (pH\,2), the agglomeration is significant, indicating that van der Waals forces overcome the charge stabilization\cite{Overbeek1977}. This and capillary forces lead to particle deformation and maximization of the particle interface as shown in the inset in Figure \ref{ScansCSPMG}a. On the other hand, the more hydrophilic MGs, characterized by their fluffy shell and long-reaching dangling ends, show a homogeneous distribution without the formation of larger aggregates independent of the particle charge. Furthermore, the MGs maintain their spherical shape on a lateral scale (inset Figure \ref{ScansCSPMG}c), and no compression is observed at the particle-particle interface.\\

\subsection{Desorption}

\begin{figure*}[h]
                 \centering
                 \includegraphics[scale=0.8]{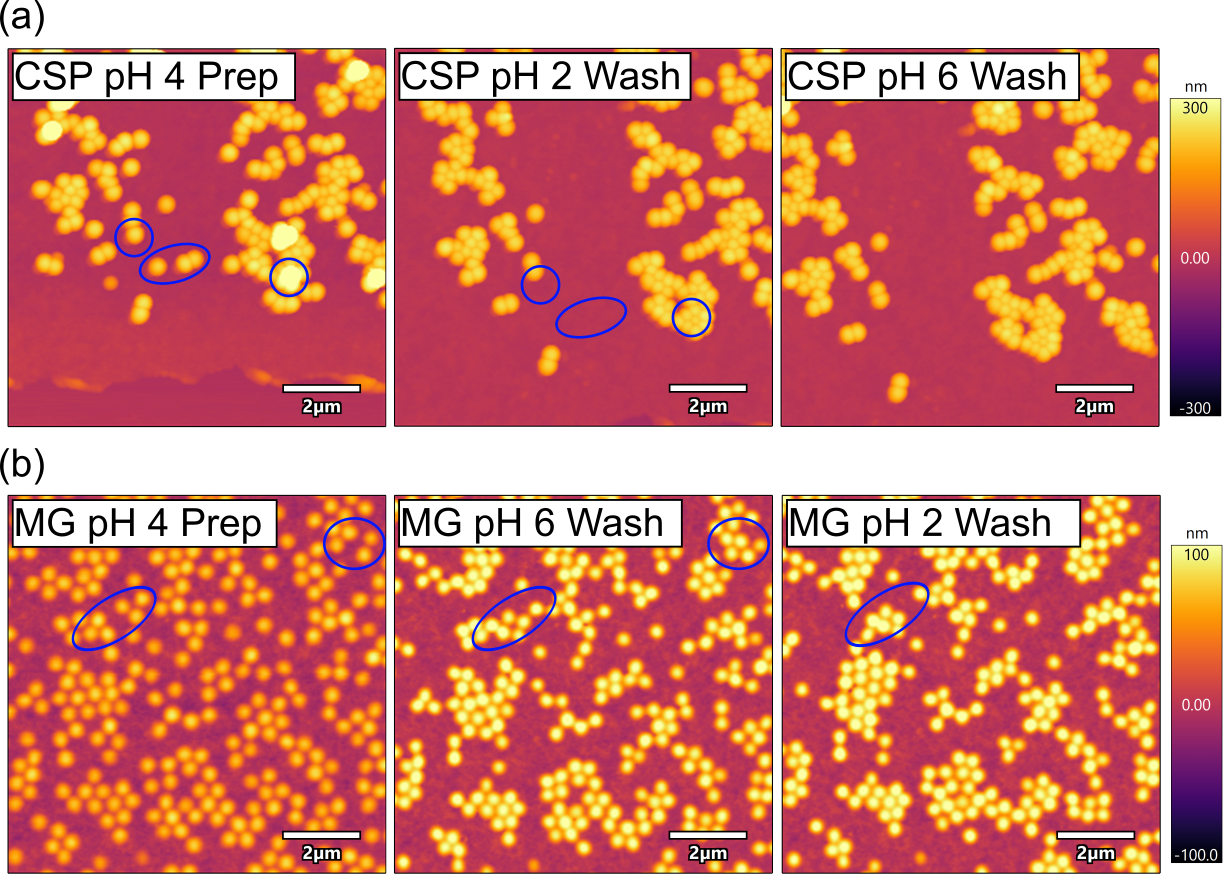}
                 \caption{The desorption of the CSP (a) and the MG (b) studied by various AFM measurements with rinsing steps in between the different topography images. All images have a size of 10x\SI{10}{\micro m} and a height scale of \SI{600}{nm} (CSP) or \SI{200}{nm} (MG). The image on the left each corresponds to the sample/composite after preparation at pH\,4. The composite film are then rinsed at pH\,2 and pH\,6 in the order listed. The desorption or movement of particles are exemplary shown by the blue markers.}
                 \label{Desorption}
\end{figure*}

First, CSPs and MGs were adsorbed at pH 4 on CMSs (Figure \ref{Desorption}). The composites are then rinsed for \SI{5}{min} at pH\,2 and then at pH\,6 for the CSP, for the MG first at pH\,6 and then at pH\,2. We choose this order, as the first rinsing step should not include a charge reversal of the CSP or charge neutralization of the MG for better comparison. In between the rinsing steps, the topography of the composite films is characterized, always choosing the same spot on the sample and the movement and desorption of the particles monitored. The composite for the CSP at pH\,4, similar to Figure \ref{ScansCSPMG}a, shows a high particle number density with some agglomeration. Rinsing the composite film with pH\,2 solution leads to the desorption of single particles and particles adsorbed in a second layer, as shown by the blue markers. In some cases, especially when the particles are closer to larger agglomerates, they do not desorb but form a close packing with the aggregates. Larger agglomerates do not seem to be influenced by the desorption. Subsequent rinsing with pH\,6 leads to no further desorption, even for single CSPs. The composite for MGs is also prepared at pH\,4 but rinsed first with pH\,6 solution. The comparison to pH\,4 shows no desorption, not even of single particles. Nevertheless, a movement is observed for many particles (approximately \SI{20}{\%}, evaluated by superposition of the images), leading to a higher packing. An overall increase in height and closer packing is found over the whole sample when rinsing with pH\,6 solution. Subsequent rinsing with pH\,2 leads to some particle movement, which mostly leads to a higher packing with no observable desorption of single particles.\\

Desorption measurements show that both CSPs and MGs exhibit irreversible adsorption. Changing the pH to uncharged conditions (pH\,6 for CSP, pH\,2 for MG) does not lead to significant desorption of either composite system. The van der Waals forces at the particle/CMS interface are strong enough that a charge reversal (pH\,6 for the CSP) or neutralization does not lead to a desorption of the particles. However, the MGs show increased agglomeration after the subsequent rinsing steps, which can be attributed to capillary forces during of the drying of the composite film\cite{Denkov1992,Schmidt2008}.\\

\subsection{Adhesion}
\begin{figure}[]
                 \centering
                 \includegraphics[scale=0.25]{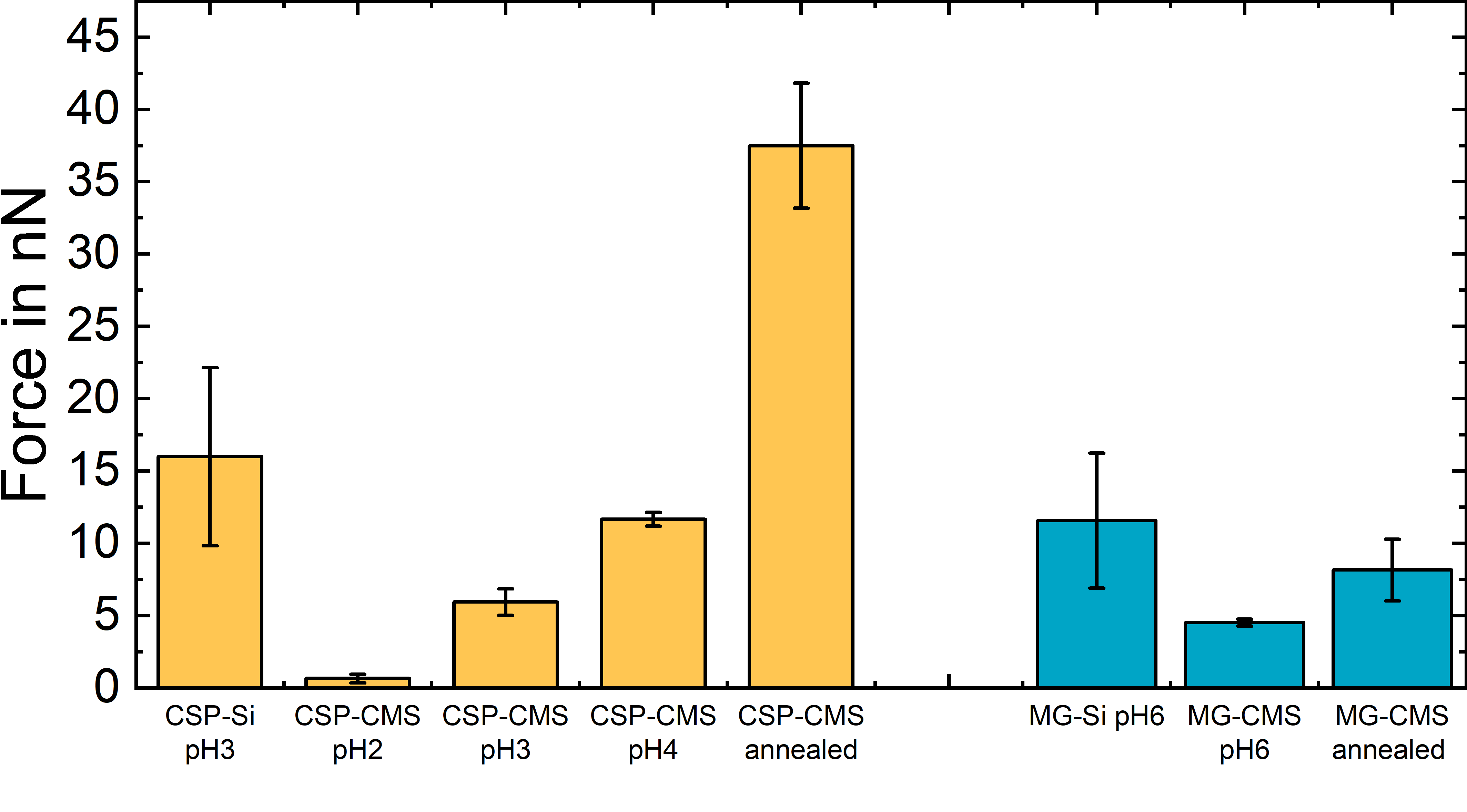}
                 \caption{Results of the pushing particles measurements for the CSP composites (yellow) and the MG composites (blue) in dependence of pH and substrate. The annealed samples were prepared and measured at pH\,3 for the CSP composite and pH\,6 for the MG composite. The results were evaluated by calculating the arithmetic mean of the number of particles moved at each force.}
                 \label{PushResults}
\end{figure}

To measure the adhesion forces of the CSPs and MGs to the CMS, pushing particle measurements were carried out at pH\,2 to pH\,4 for CSPs and at pH\,6 for MGs. These measurements are compared to those with CSPs and MGs adsorbed on a silicon wafer. We chose these samples to study the impact of charge on the adhesion (CSP row) and the impact of structure/dangling ends (comparison MG and CSP). By peak force AFM measurements, composite films were scanned with increasing force and the movement or desorption of particles monitored. An example is shown in Figure S2. The results of all measurements are summarized in Figure \ref{PushResults}. The distribution is calculated based on monitoring the number of particles moved at each force and carrying out the measurement until all particles have been removed from the surface.\\

The pushing particle measurements with CSPs show an increasing adhesion force with pH. For pH\,2, where the CMS is neutral or even charged positively, the particles already detach at a low force of \SI{0.7(3)}{nN}. Adhesion increases with increasing pH to \SI{11.4(5)}{nN} at pH\,4. In comparison, CSPs on a silicon wafer at pH\,3 adhere up to \SI{16.0(62)}{nN}. While the distribution of the forces necessary to move the particles on the CMS is narrow, the values on the silicon wafer cover a higher range. MGs on CMS at pH\,6 adhere at a force up to \SI{4.5(2)}{nN}. In comparison, MGs stick better to the silicon wafer with forces of \SI{11.6(47)}{nN} necessary to move the particles. It can again be observed that the adsorption on the silicon wafer leads to a significantly broader force distribution compared to the adsorption on the CMS.\\

The measurement shows that the charges of the composite system play a dominant role in particle-CMS adhesion. Increasing the charge density of the composite from pH\,2 (CMS is uncharged) to pH\,3 (both CMS and CSP are charged), increases the electrostatic interaction and leads to an increase in adhesion. The same is observable when further increasing the pH to pH\,4 (the charge of CMS significantly increases) leading to strong adhesion. Despite the similarity in electrostatic behavior between MGs at pH\,6 and CSPs at pH\,3, and the larger interface of MGs compared to CSPs (Figure \ref{Crosssection}), the CSPs exhibit stronger adhesion. This difference can be attributed to the interaction between MGs and CMS occurring at individual chains with a high water content in between rather than a compact interface. This result in a higher van der Waals interaction for the CSPs than for the MGs at the interface. Furthermore, CSPs predominantly carry charges on the particle surface, while MGs are characterized by a more distributed charge throughout the particles.\\

For both particles adsorbed on an etched silicon wafer (CSP at pH\,3, MG at pH\,6), the adhesion force is significantly stronger compared to the respective composites with the CMS. The silicon oxide layer on the wafer has a negative charge, which increases with increasing pH.\cite{Blumenstein2016} The stronger adhesion of the CSPs to the \ce{SiO2}-layer can be explained by the stronger electrostatic force resulting from the higher negative charge on the silicon wafer compared to the CMS (\SI{-60}{mV} for \ce{SiO2} and approx. \SI{0}{mV} for CMS at pH\,3). At pH\,6, the silicon wafer and CMS exhibit similar charges, and the increased adhesion of MGs to the \ce{SiO2}-wafer can be attributed to a higher surface energy of the wafer, an enhanced contact area between MG and the wafer, or increased hydrogen bonding with the silicon oxide \cite{Wu1999} compared to the CMS.\\

The measurement of the composites of CSPs at pH\,3 and MGs at pH\,6 were repeated on "annealed" composites that underwent 10 swelling/drying cycles in pH\,3 water. In both cases, the treated composites showed higher adhesion forces than the respective untreated samples. For the treated CSP composite a force of \SI{37.5(43)}{nN} is measured, six times higher than for the untreated composite. For the MG composite a force of \SI{8.2(21)}{nN} is measured, twice as high as for the untreated MG composite. Similar to the measurements on a silicon wafer, these measurements give a broader force distribution.\\

\begin{figure}[h]
                 \centering
                 \includegraphics[scale=0.8]{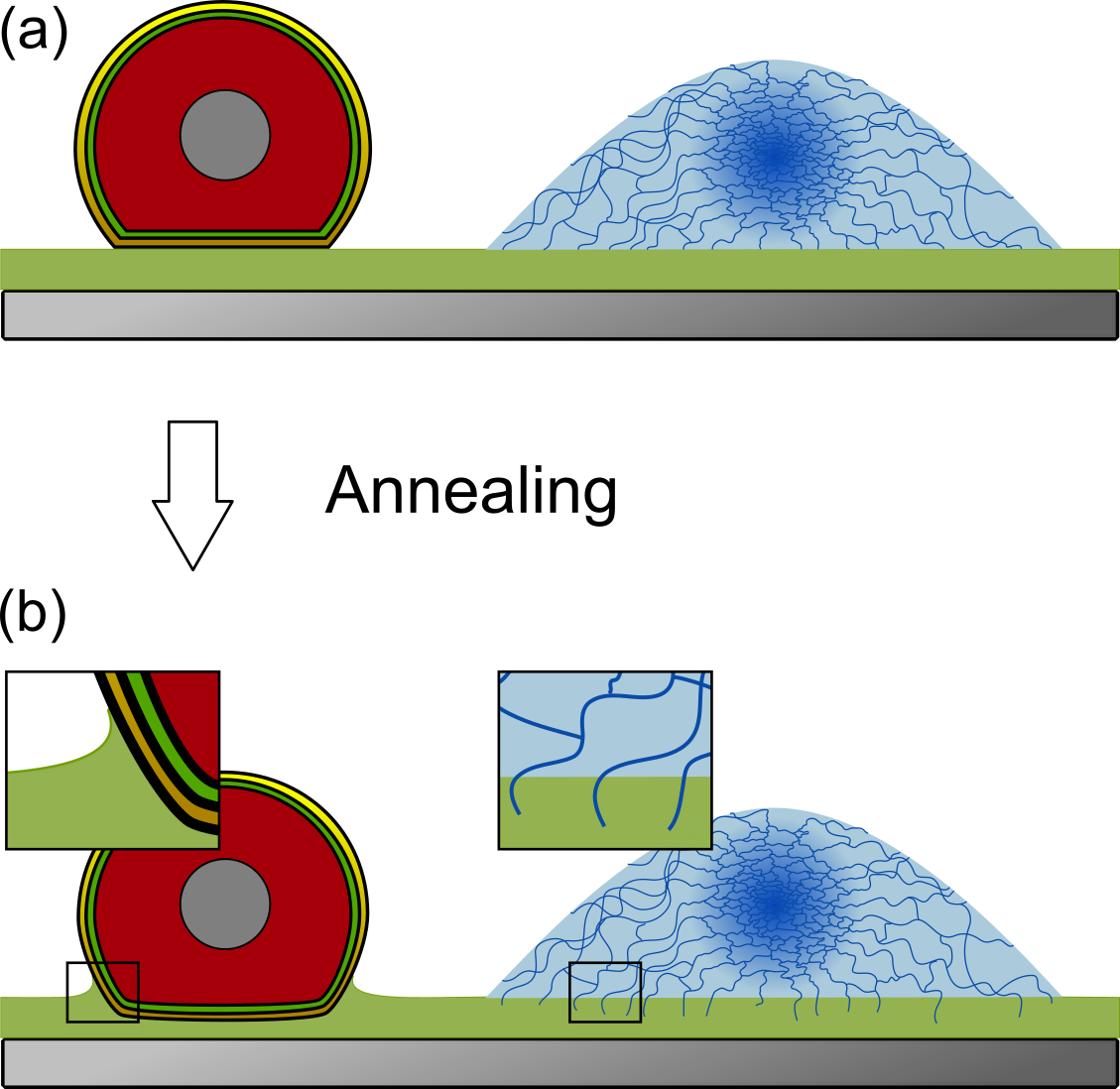}
                 \caption{Scheme of the adsorbed CSP (left) and MG (right) on the CMS. When the untreated composite (a) undergoes several swelling and drying cycles, which we describe as an annealing (b), the interface with the CSP increases, while the loose chains of the MGs penetrate into the CMS.}
                 \label{DiscScheme}
\end{figure}

The untreated composites exhibited comparable adhesion, whereas the annealed CSP composite displayed a fivefold increase in adhesion force, whereas the MG only doubled in force. This difference can be attributed to the dense structure of the CSP, which maintains an unchanged CSP/CMS interface. The increased adhesion can only result from an increase in contact area \cite{Pollock1978}, as depicted in Figure \ref{DiscScheme}. The CMS undergoes swelling, doubling its thickness with a high water uptake. This swelling likely causes an irreversible increase in the CSP/CMS interface. In contrast, the interface of MG with CMS is already maximized, and therefore, CMS swelling does not impact the soft MG particles, in the sense that they are enveloped by CMS. In comparison to the CSP, the dangling ends of the MG resemble linear polymer chains, which can interact with the CMS. Horvarth et al. showed that cationic polyelectrolytes are able to diffuse into the cellulose fiber wall\cite{Horvath2008}, resembling the mobility of polyelectrolyte chains in polyelectrolyte multilayers \cite{Lux2021,Nazaran2007}. Backes et al. showed a interdigitation of the dangling ends of a negatively charged P(NIPAM-co-AAc) MG and polymer film from poly(allylamine hydrochloride).\cite{Backes2017a} The different impact of annealing on the adhesion of CSPs and MGs can therefore be explained by these distinct mechanisms at play. In retrospect, annealing can already take place during  the rinsing steps, as shown in chapter \ref{Desorption}.\\

\section{Summary and conclusion}
\label{conclusion}
In this study, we prepared a well defined cellulose model surface (CMS) with a low roughness and studied the interaction with polymeric particles to understand the functionalization of paper materials. The used particles were a densely crosslinked, spherical core-shell particle (CSP) and a loosely crosslinked PNIPAM microgel (MG), both positively charged. We showed that the charges of both the particles and the CMS and the electrostatic interaction control the adsorption. Due to the structure of the particles - the dense interface of the CSPs and the dangling ends of the MGs - the adsorption is irreversible. No desorption takes place, even after a pH induced charge reversal of the particles. However, to better understand the interaction the structural morphology of the interface between particles and CMS had to be considered. This is evident as the total contact area of the particle and the electrostatic interaction do not fully explain the adhesion as shown by the pushing particles measurements. Due to the stiffness of the CSP and the maintaining of the spherical structure, the interaction of the CSP only takes place at the outer surface of the CMS. The softer MG deforms for an interface maximization and the dangling ends behave more like linear chains, being able to penetrate into the CMS. The interface is tunable as shown by an annealing procedure, which allows the increase of the adhesion for both the CSP and the MG composites. The CSPs adhere stronger than the MG which might be due to their higher surface charge density and denser structures. The tunable adhesion of the particles has the potential to create papers with manifold applications. Comparing the morphologies and the resulting properties such as deformability and adhesion, shows that particles with the combined properties of CSP and MG show prospect for application in flexible papers. Particles with the deformability similar to the one of the MGs presented here, but the charge density and adhesion of the CSP would have the potential to create flexible and mechanically stable papers. A MG, copolymerized with a positively charged monomer such as allylamine, and a high crossslinker density could fulfill the desired properties of such a particle. By expanding the understanding of particle-CMS interactions in this research, the development of advanced materials and applications in various fields can be achieved.
\newpage

\begin{acknowledgement}
The authors thank the Deutsche Forschungsgemeinschaft (DFG) under the grant PAK 962-2, subprojects
KL1165/28-2 (Project No 405629430) and STA 1026/8-2 (Project No 405549611), for the funding of this project.
\end{acknowledgement}

\subsection{Conflict of Interest}
The authors have no conflict of interest to declare.


\newpage

\bibliography{library}

\newpage

\begin{suppinfo}
\setcounter{figure}{0} 
\renewcommand{\thefigure}{S\arabic{figure}}

\pagenumbering{Roman}

\begin{figure}[]
                 \centering
                 \includegraphics[scale=1]{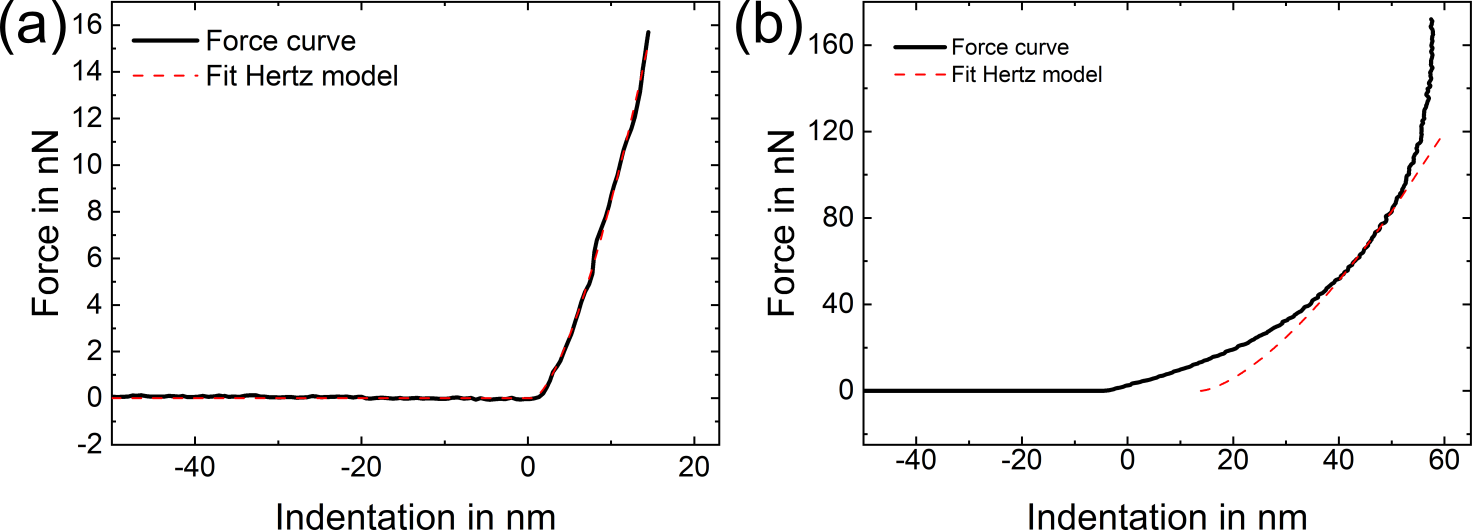}
                 \caption{Exemplary force curves and fitting at (a) low indentations and (b) high indentation using an offset in the Hertz model.}
                 \label{SI:ExFC}
\end{figure}

\begin{figure}[]
                 \centering
                 \includegraphics[scale=0.8]{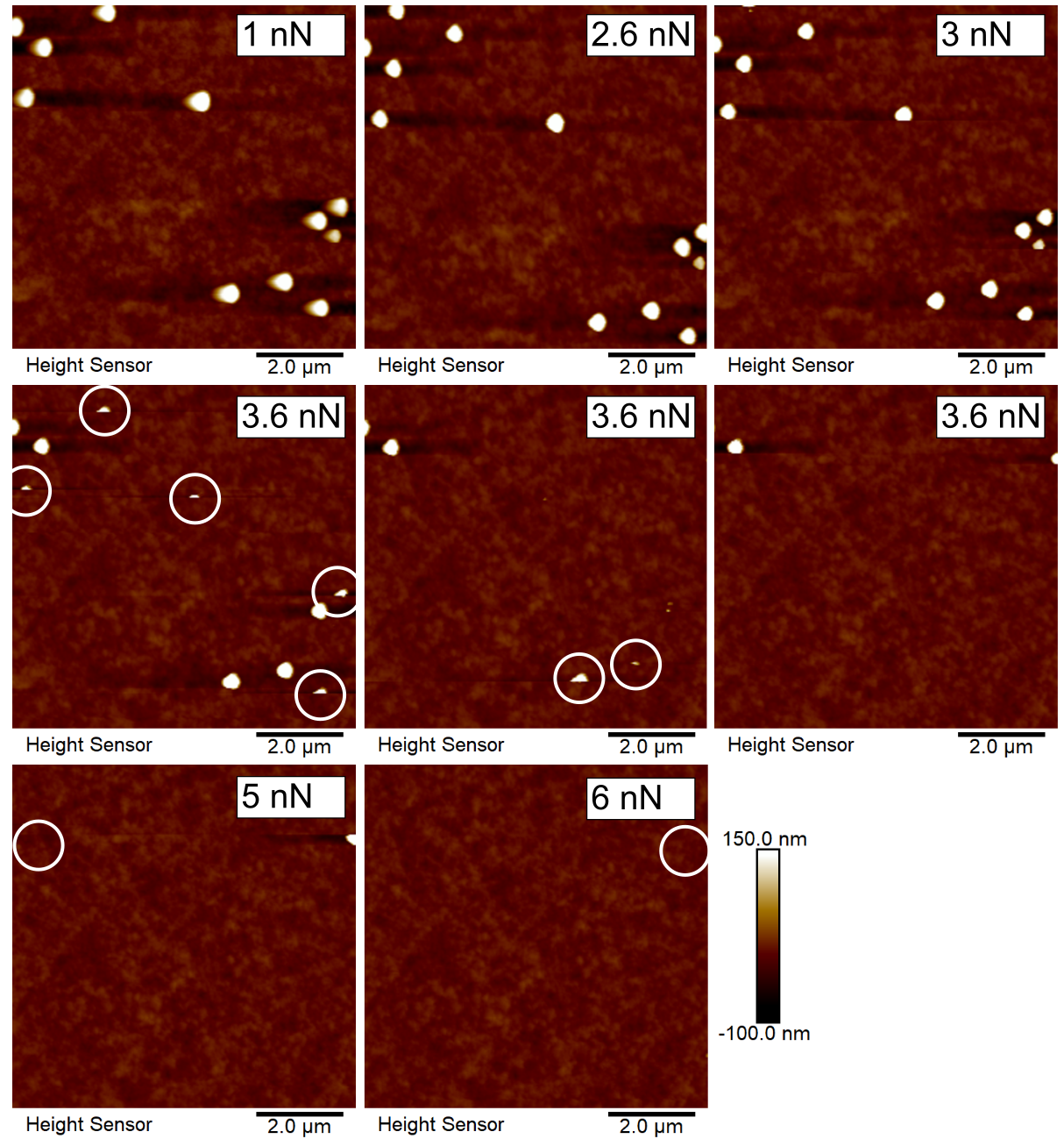}
                 \caption{Exemplary results of the pushing particles measurement for the CSP on CMS at pH\,3. All images have a size of 8\,x\,\SI{8}{\micro m} and a height scale of \SI{250}{nm}. The movement of the particles is monitored and shown here with the white markers.}
                 \label{SI:ExPushRes}
\end{figure}

\begin{figure}[]
                 \centering
                 \includegraphics[scale=0.8]{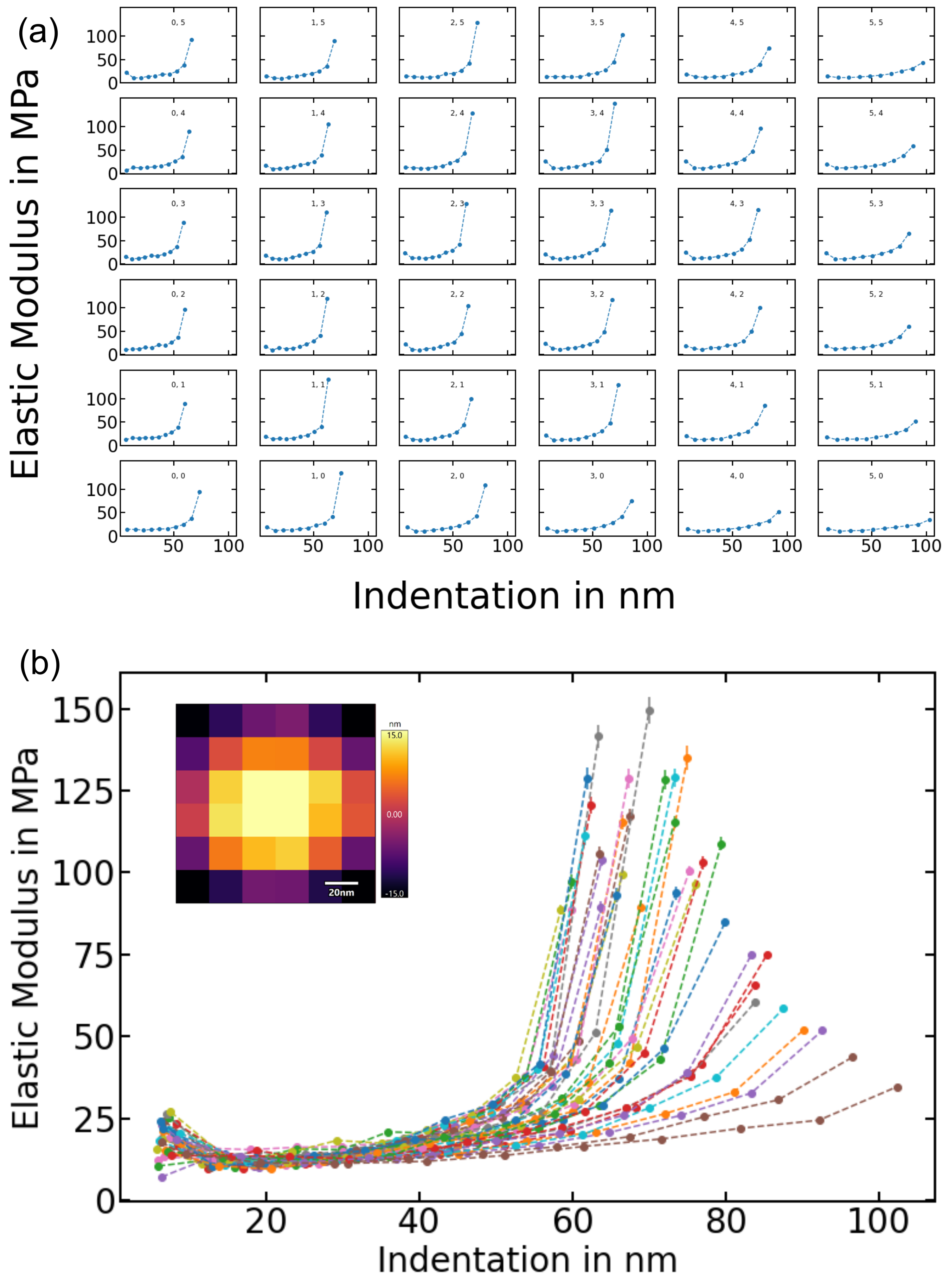}
                 \caption{Indentation measurements of the CSP at varying indentation depth. (a) shows the \textit{E} modulus in dependence of the indentation depth for specific position measured at the center of the particle. The graphs are ordered to match the height map shown in (b) which has a size of 100\,x\,\SI{100}{nm}. (b) also shows all the \textit{E} moduli curves superimposed for easier comparison.}
                 \label{SI:Indepth_CSP}
\end{figure}

\begin{figure}[]
                 \centering
                 \includegraphics[scale=0.8]{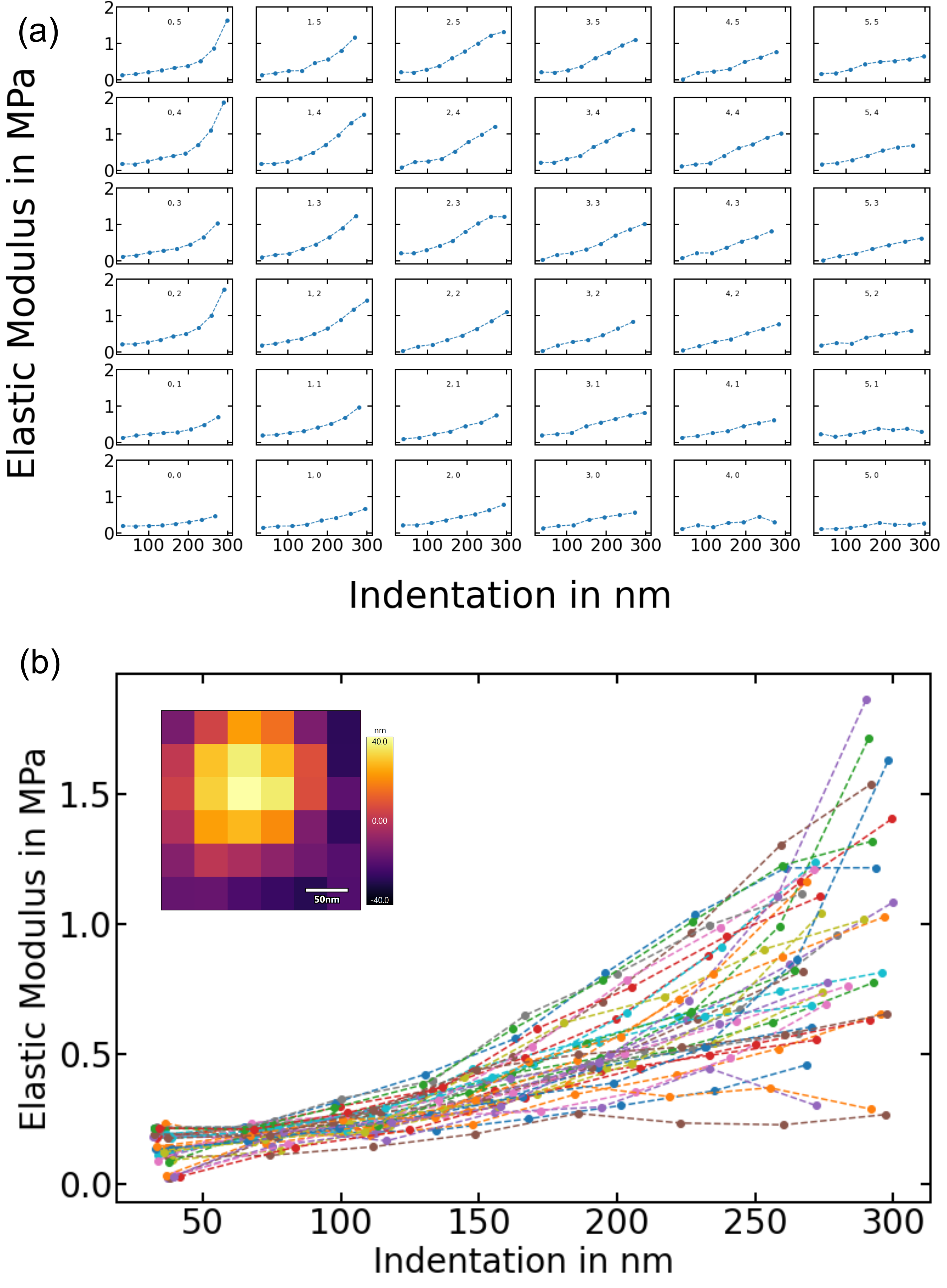}
                 \caption{Indentation measurements of the MG at varying indentation depth. (a) shows the \textit{E} modulus in dependence of the indentation depth for specific position measured at the center of the particle. The graphs are ordered to match the height map shown in (b) which has a size of 200\,x\,\SI{200}{nm}. (b) also shows all the \textit{E} moduli curves superimposed for easier comparison.}
                 \label{SI:Indepth_MG}
\end{figure}

\end{suppinfo}

\end{document}